# Patterns and Pathways: Applying Social Network Analysis to Understand User Behavior in the Tourism Industry Websites

Mehrdad Maghsoudi [a,*], Saeid Aliakbar [b] , AmirMahdi Mohammadi [c]

[a] *Department of Industrial and Information Management, Faculty of Management and Accounting, Shahid Beheshti University, Tehran, Iran*

[b] *Department of Industrial Engineering, Engineering Faculty of Damghan University, Semnan Province, Iran*

[c] *Department of Mechanical Engineering, Faculty of Engineering, K. N.Toosi University, Tehran, Iran*

**Abstract**

The contemporary tourism landscape is undergoing rapid digitization, necessitating a nuanced comprehension of online user behavior to guide data-driven decision-making. This research bridges an existing gap by investigating the tourism website ecosystem through social network analysis. It focuses specifically on inter-website communication patterns based on user navigation. Data mining facilitates the identification of 162 core Iranian tourism websites, which are visualized as an interconnected network with websites as nodes and user transitions as weighted directed edges. By implementing community detection, eight key clusters are discerned, encompassing domains like ticket/tour bookings, accommodations, location services, and cuisine. Further analysis of inter-community relationships reveals website groupings frequently accessed together by users, highlighting complementary services sought during travel planning. The research derives invaluable insights into user preferences and information propagation within the tourism ecosystem. The methodology and findings contribute original perspectives to academia while offering pragmatic strategic recommendations to industry stakeholders like service providers, investors, and policymakers. This pioneering exploration of latent user behavior patterns advances comprehension of the evolving digital tourism landscape in Iran. It contributes pathways toward a sustainable future vision of the ecosystem, guiding stakeholders in targeted decision-making based on empirical evidence derived from social network analysis of websites and consumption patterns. The innovative methodology expands the toolkit for data-driven tourism research within academia.

---

[*] Corresponding author. e-mail: M_Maghsoudi@sbu.ac.ir.

## 1. Introduction

The contemporary tourism industry is undergoing a transformation driven by factors such as the rise in global travel, continuous infrastructure improvements, and the pervasive penetration of digital technologies (Christou et al., 2023). This exponential growth has been further amplified by the digitization of daily activities and the increasing availability of online services, positioning the tourism sector as a fertile ground for such emerging trends (Büyüközkan & Ergün, 2011) Consequently, businesses within the tourism industry are persistently striving to understand user behavior, identify vital needs, and capitalize on unique opportunities to enhance profitability and competitiveness (Ozdemir et al., 2023).

The fundamental significance of comprehending user behavior as a determinant of success for tourism service providers has come to the forefront (Kozak & Buhalis, 2019). Understanding user behavior is pivotal in not only attracting audiences but also in creating an environment and services that bolster satisfaction and loyalty (Pasca et al., 2021). Hence, unraveling the intricacies of this ecosystem and deciphering the relationships interconnecting its diverse components has become crucial for all stakeholders in this industry, including service providers, investors, and policymakers (Yanes et al., 2019).

Despite the paramount importance of exploring and analyzing this multifaceted ecosystem, a noticeable deficiency in comprehensive research within this domain exists. This scarcity poses significant challenges for stakeholders in the tourism industry who are aiming for data-driven decision-making and alignment with the evolving landscape (Dolnicar et al., 2014; Polese et al., 2022; Zhao & Yu, 2021). This research seeks to bridge this gap by delving into the user behavior landscape and conducting an analysis of the Persian-language website ecosystem. To achieve this goal, the study focuses on the consumption behavior of Iranian users of tourism services and employs social network analysis (SNA) to uncover and analyze complex communication patterns within the Iranian tourism website ecosystem.

Social Network Analysis (SNA) is a powerful and illuminating method that enables deep exploration of mutual connections and communication patterns among various network elements (HabibAgahi et al., 2022). In the realm of the tourism industry, SNA allows us to visualize relationships among tourism websites, users, service providers, and other stakeholders. Analyzing social networks by scrutinizing their structure and information flow provides valuable insights into the dissemination of information, preferences, and decisions within the ecosystem (Maghsoudi, Shokouhyar, et al., 2023).

By examining the nexus between user behavior and the tourism ecosystem through SNA, this study aspires to contribute to the advancement of the tourism industry and offer practical insights to industry stakeholders. Through an extensive exploration of user behavior and communication patterns, this study aims to enrich the understanding of the dynamic and evolving landscape of the digital era tourism industry, paving the way for a more sustainable and innovative future.

The structure of this article is as follows. Chapter 2 provides a review of the research literature, social network analysis, and the endeavors of other researchers in the data-driven analysis of the tourism industry. The research methodology, detailed in Chapter 3, outlines the systematic approach adopted to study user behavior and implement social network analysis. Chapter 4 presents the research findings, offering valuable insights into behavioral patterns within the Iranian tourism ecosystem derived from SNA results. Finally, in Chapter 5, by combining the key contributions of the study and presenting policy recommendations based on the findings, the research will be concluded.

## 2. Literature Review

In the current context of the tourism industry, its dynamic nature and vast scope are apparent (Rwigema, 2021). This sector, spanning activities related to hospitality, transportation, entertainment, and more, has emerged as a dominant global force across various economic indicators, encompassing gross production, value addition, employment, investment, and tax contributions (Beladi et al., 2017; Gudkov & Dedkova, 2020; Ridzuan & Abd Rahman, 2021). Notably, the industry is composed of multifaceted ventures that provide a range of services, including lodging, food, beverages, transportation, and recreational activities (Arroyo et al., 2021; Bodhanwala & Bodhanwala, 2022; Jeaheng & Han, 2020; Martín Martín et al., 2021; Okumus, 2021). This complexity of the tourism industry underscores the need for stakeholders to comprehend its distinctive attributes, ensuring informed decision-making (Gursoy et al., 2022). These distinctive aspects encompass (Holloway & Humphreys, 2022):

- Tourism transcends mere traditionalism; it consists of an amalgamation of diverse personal ventures spanning hospitality, transportation, entertainment, and other spheres.
- Its essence lies in providing service, emphasizing the significance of service quality.
- Competitiveness is inherent, compelling enterprises to foster innovation to retain a competitive edge.
- The industry constitutes a complex, multifaceted ecosystem involving a spectrum of stakeholders such as service providers, investors, policymakers, and tourists.

Recognizing the intricacies of user behavior holds substantial significance within the tourism sector (Carvalho & Alves, 2023). By scrutinizing user behavior patterns, businesses can cultivate audience engagement, and design platforms fostering contentment and loyalty (Gupta, 2019; Maghsoudi & Nezafati, 2023). The scrutiny of user behavior patterns allows the identification of critical needs and the exploration of untapped prospects to bolster profitability and competitiveness (Ozdemir et al., 2023; Zhang et al., 2019). Beyond businesses, policymakers and investors in the tourism realm equally benefit from user behavior analysis, as it facilitates data-driven decision-making in alignment with the evolving landscape (Chen et al., 2021; Kwok, 2023).

### 2.1. Social Network Analysis

Social Network Analysis (SNA) encompasses an array of methodologies and techniques with a primary focus on constructing a community based on interrelationships among actors and dissecting structures that emerge from the recurrence of these relationships (Maghsoudi, Shokouhyar, et al., 2023). The methodology and theoretical underpinning of SNA offer a comprehensive, top-down viewpoint into the dynamics among individuals, elements, and entities in a network (M. A. M. A. Kermani et al., 2022). Social networks are harnessed to unveil the relational patterns among network participants, grounded in the foundational premise that scrutinizing connections between entities yields all-encompassing insights into the network (Maghsoudi & Shumaly; Zohdi et al., 2022). A key differentiation of SNA from other methodologies lies in its emphasis on the interactions between elements rather than exclusively highlighting the distinct attributes of each individual element (HabibAgahi et al., 2022; Jalilvand Khosravi et al., 2022). In the context of network analysis, the interplay among members serves as a fundamental principle, wherein a social network analyst strives to comprehend how entities come into being and establish connections within a network (HabibAgahi et al., 2022). Social networks demonstrate distinctive characteristics arising from the relational configurations between their constituent elements, which do not adhere strictly to regularity or randomness. Visualizing data often allows researchers to uncover patterns that might otherwise remain concealed (Maghsoudi, Shokouhyar, et al., 2023).

One avenue for conducting social network analysis involves employing graph theory (Maghsoudi, Jalilvand Khosravi, et al., 2023). A network comprises two core components: 1) Nodes, representing discrete points within a graph, and 2) Edges or links, connecting pairs of nodes (Jalilvand Khosravi et al., 2022). Consequently, by extracting the relational structure and communication patterns among these actors and employing graph-theoretical models in mathematics to illustrate these associations using lines and active network participants, the target network is simulated in the form of a graph or an array of interconnected lines and nodes (M. A. Kermani et al., 2022). In the analysis of networks, elements with robust internal connections and limited external connections within the social network structure are labeled as communities. These communities can be perceived as relatively autonomous constituents within the social network framework. In the realm of social networks, a community denotes a subgraph of nodes that are closely interlinked internally and less connected externally (Maghsoudi, Shokouhyar, et al., 2023). The subsequent list encapsulates the most prominent attributes and benchmarks of social networks:

- Nodes: Individuals, members, or entities existing within a network.
- Edges: Links between nodes in the network.
- Size: Reflects the count of nodes in the network.
- Degree: Quantifies the number of edges in a network.
- Community: A subset of nodes exhibiting heightened internal interconnections compared to the broader network (HabibAgahi et al., 2022; Zohdi et al., 2022). Tools for community detection facilitate the identification of the most active entities within a network.

### 2.2. Communities in Social Network Analysis

In a broader context, a community in the real world represents a collection of individuals who share common economic, social, or political interests and often coexist in proximity. When users establish their presence on social media platforms and engage in interactions, a virtual community emerges. In essence, each community must consist of a minimum of two nodes where shared interests foster a degree of commitment (Bedi & Sharma, 2016). Communities can be characterized as clusters of nodes that exhibit a closer interconnection with each other compared to other nodes within the network. An intrinsic community is shaped by individuals in a manner that encourages greater interaction among its members than with those outside the community (Figure 1). The extent of closeness among entities within a community can be gauged through an assessment of their similarities or disparities (Maghsoudi, Shokouhyar, et al., 2023). In the realm of social networks, a community can be likened to a distinctive cluster within the broader network fabric (Jalilvand Khosravi et al., 2022).

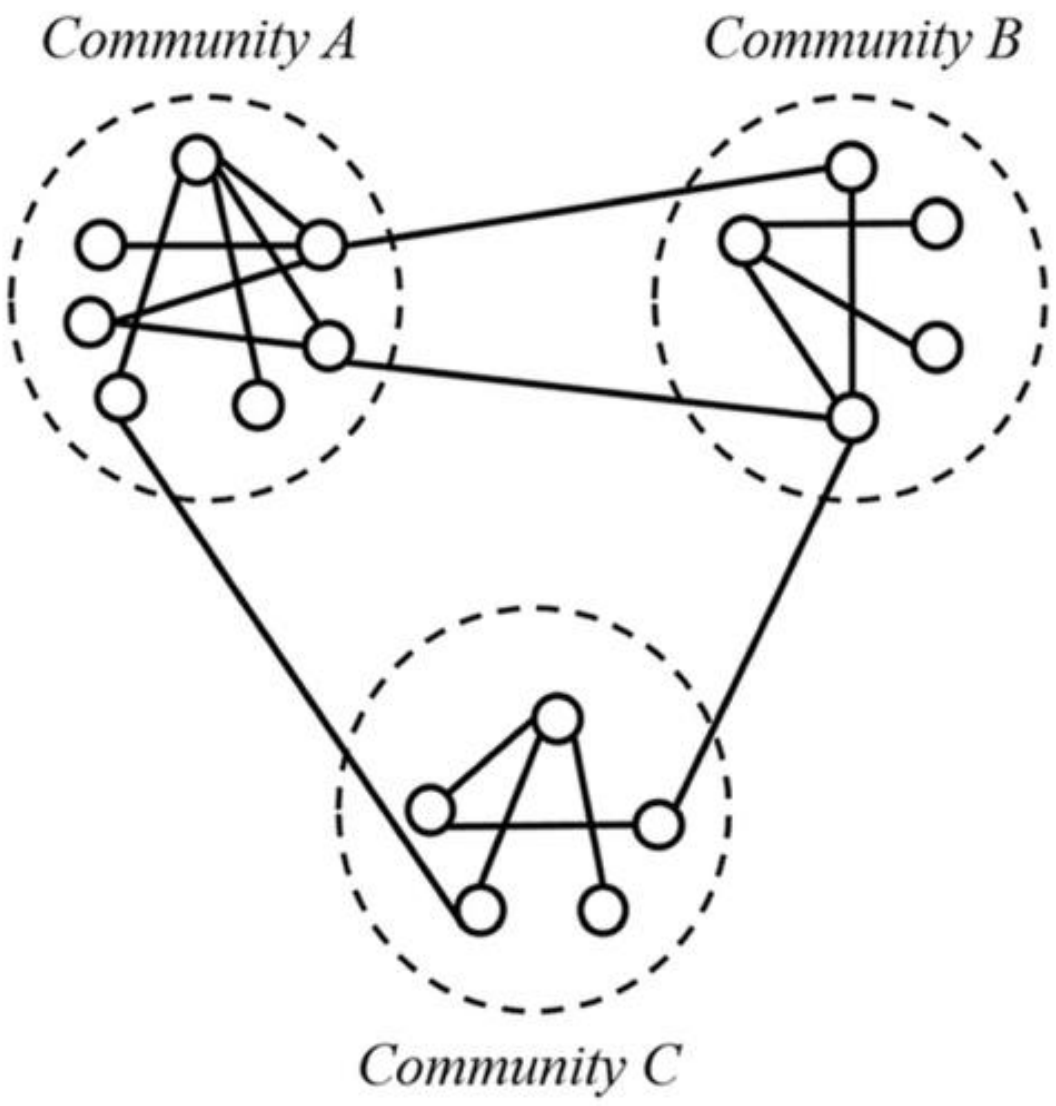

*Figure 1: Communities in the network*

The identification of communities holds a significant role in the recognition of cohesive structures utilizing clustering methodologies. A visual depiction of this procedural occurrence is offered in Figure 2. A community is characterized as a cluster of closely interconnected nodes, exhibiting stronger internal connections than those to nodes outside the community. These community formations are of particular value to network analysts, as they unveil interaction patterns and interconnected subgroups within the network (Zafarani et al., 2014).

Modularity, as introduced by Newman (Newman, 2006), stands as one of the prevalent metrics employed to evaluate the efficacy of social network clustering. Modularity serves as an indicator of the extent of clustering within a network, where higher values correlate with more robust interconnections among linked nodes.

Consider a weighted network comprising n nodes. The algorithm's initial step treats each node as an independent community, resulting in the establishment of multiple communities initially. Subsequently, for each node i, the algorithm identifies the adjacent community j that maximizes the modularity score when node i transitions from its current community to community j. This reallocation only transpires if it enhances the overall modularity. Alternatively, node i remains within its original community. This iterative procedure continues for all nodes until no further alterations occur, culminating in an optimal local state where modularity can no longer be improved, signifying the conclusion of the initial phase.

In the algorithm's subsequent phase, smaller communities are amalgamated to forge larger entities. The objective of this stage is to optimize the modularity score by merging compatible communities. This dual-phase process persists until community configurations remain unaltered, and the modularity score reaches its peak state.

Figure 2 serves to illustrate the algorithm's operational mechanism. As depicted in the image, the algorithm effectively discerns four distinct communities after the initial phase, converging towards a local optimum. In the second phase, the algorithm endeavors to unite these four communities into two larger entities, thereby attaining the maximum modularity score and finalizing the process.

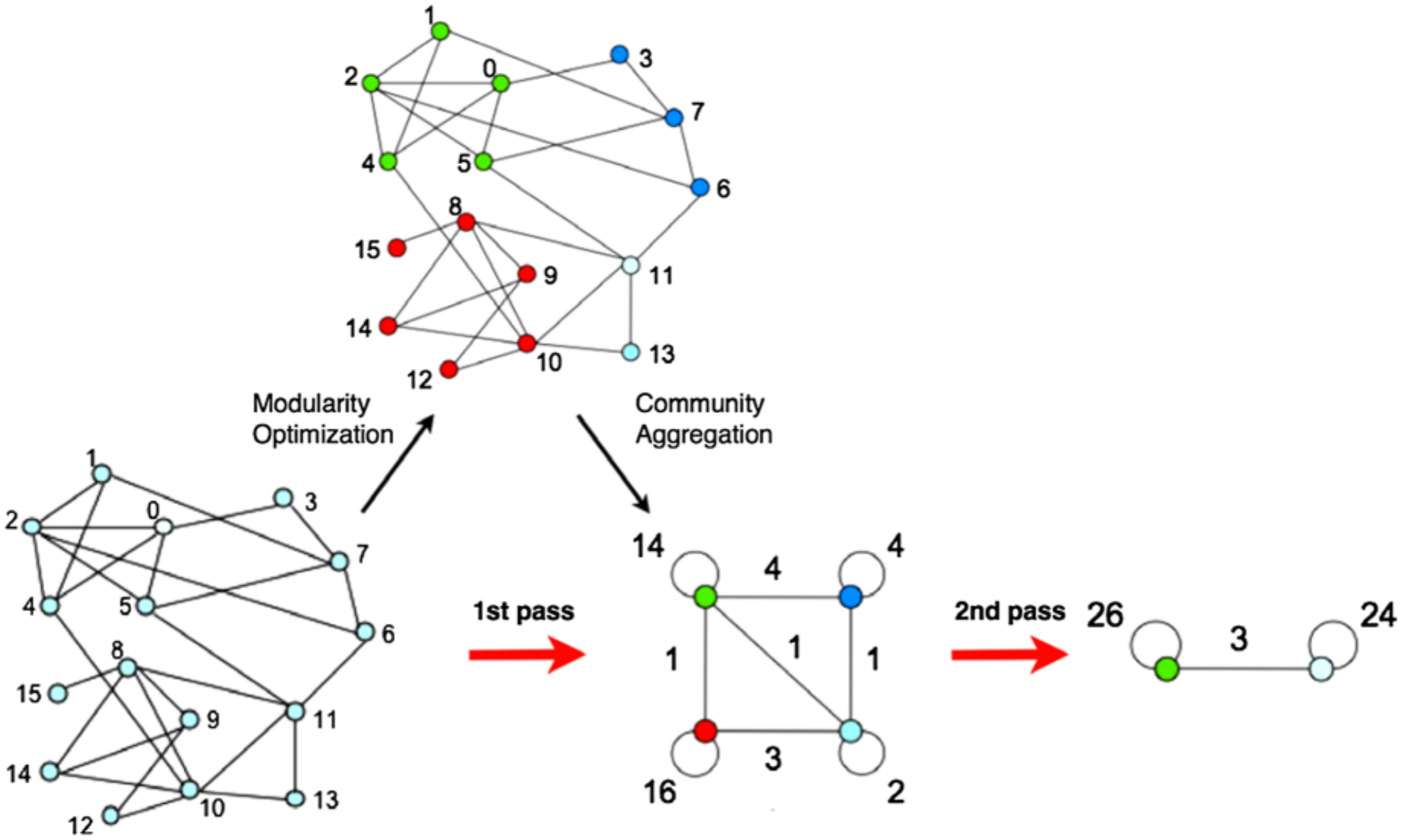

*Figure 2: The stages of community detection*

### 2.3. Related works:

This section examines a series of studies covering different facets of the travel and tourism sector. These investigations illuminate the application of cutting-edge technologies, data-driven approaches, and analytical techniques in this industry. The insights from these works significantly enhance our understanding of how artificial intelligence, big data, machine learning, and statistical analysis are impacting tourism. Specifically, they are leading to advancements in personalization, customer value, marketing research, and pattern analysis.

(Gupta et al., 2023) present research exploring the potential of AI-powered facial recognition to augment the value proposition in travel and tourism. Through interviews and thematic analysis, they identify four main themes: personalization, data-driven services, security/safety, and integrated payments. This highlights how AI-based facial recognition can improve understanding of travelers' needs, optimize service offerings, and streamline value-based services. Drawing on data-driven insights like customized trip planning and efficient payment systems, this study delivers valuable knowledge grounded in organizational information processing theory.

(Weaver, 2021) tackles challenges arising from big data analysis in tourism. Through reviewing tourism journals, they highlight concerns about declining individual influence as the industry increasingly relies on impersonal mass data. This analytics crisis creates tensions, potentially impacting tourist and industry stakeholder dynamics, and influencing capital accumulation. The paper advocates for a balanced, human-centric approach to data analysis that affirms the distinct, creative human actions shaping tourism.

In the 2021 study "Tourism demand forecasting with online news data mining," conducted (Park et al., 2021), the focus is on utilizing news discourse to predict tourist arrivals in the context of Hong Kong. The research employs structural topic modeling to identify key topics and their meanings related to tourism demand. It then examines the impact of these extracted news topics on tourist arrivals to forecast tourism demand using a method that combines seasonal autoregressive integrated moving averages with the selected news topic variables. The study affirms that incorporating news data significantly enhances forecasting accuracy. Interestingly, the forecasting model based on news topics also demonstrated superior performance during instances of local-level social unrest in the destination. These findings contribute to the field of tourism demand forecasting by integrating discourse analysis and provide a tool for tourism destinations to navigate the effects of news media on various externalities.

The (Shapoval et al., 2018) study explores the utilization of data mining tools, specifically decision trees, to analyze the behavior of inbound tourists in Japan to enhance future destination marketing strategies. This pioneering research employs data mining to investigate approximately 4,000 observations and reveals that visitors' likelihood of returning to Japan is primarily influenced by anticipated future experiences rather than their past visit experiences. Notably, factors like visiting hot springs and enjoying natural landscapes play a significant role in motivating future returns. The data mining approach minimizes researcher subjectivity and enables the identification of visitor patterns within extensive datasets. This research underscores the potential of data mining as a valuable tool for governments and destination marketing organizations to develop more effective strategies for promoting tourism destinations.

(Ali Ghasemian et al., 2021) demonstrate integrating big data analytics and machine learning to uncover customer value components in Iranian tourism. Analyzing extensive online customer feedback via data clustering and association rule mining, they prove this approach enables faster, more accurate, and more comprehensive marketing research at relatively lower costs. The findings have important implications for marketing scholars and hospitality managers.

(Fararni et al., 2021) study focuses on addressing the abundance of tourist data generated through the Internet and Online Travel Agencies (OTAs). While these sources offer extensive possibilities for tourists, the sheer volume of options can overwhelm and obscure relevant results. To assist tourists in efficiently planning their trips and finding pertinent information, the study explores recommender systems. The article presents an overview of recommendation approaches employed in the tourism domain, leading to the proposal of a conceptual framework for a tourism recommender system based on a hybrid recommendation approach. This system aims to go beyond suggesting a mere list of tourist attractions tailored to individual preferences. Instead, it functions as a comprehensive trip planner, designing detailed itineraries that encompass diverse tourism resources for specific visit durations. The ultimate objective is to create a recommender system, rooted in big data technologies, artificial intelligence, and operational research, to promote tourism in the Daraa-Tafilalet region of Morocco.

(Obogo & Adedoyin, 2021) emphasize the importance of data-driven business analytics models for UK tourism, considering external impacts like COVID-19. Given tourism's major economic contribution, reliable demand forecasting is crucial. Utilizing macroeconomic data, they trial a machine learning algorithm to evaluate post-COVID industry futures, providing insights to shape local recovery strategies.

This study aims to fill the research gap on Persian tourism websites using Social Network Analysis (SNA) to examine user behavior and communication patterns. Unlike prior studies focused on AI, big data, and machine learning, this introduces a novel approach to applying SNA to map tourism website relationships. By analyzing connections and information flows, it provides valuable insights into how preferences, decisions, and knowledge diffuse to advance tourism and enable data-driven choices. The uniqueness lies in comprehensively exploring latent user behaviors, and pioneering a more sustainable, innovative future for the dynamic, digital-age tourism industry. This innovative approach seeks to enrich the understanding of Iran's tourism ecosystem and give stakeholders practical insights to ensure an accurate, holistic perspective on shaping the region's tourism future.

## 3. Methodology of the Study

This research endeavor is designed to ascertain the intricate network of interconnections among tourism industry websites. Employing a quantitative approach and leveraging the methodology of social network analysis, the study unfolds in four distinct phases. The overarching procedural roadmap can be discerned in Figure 3.

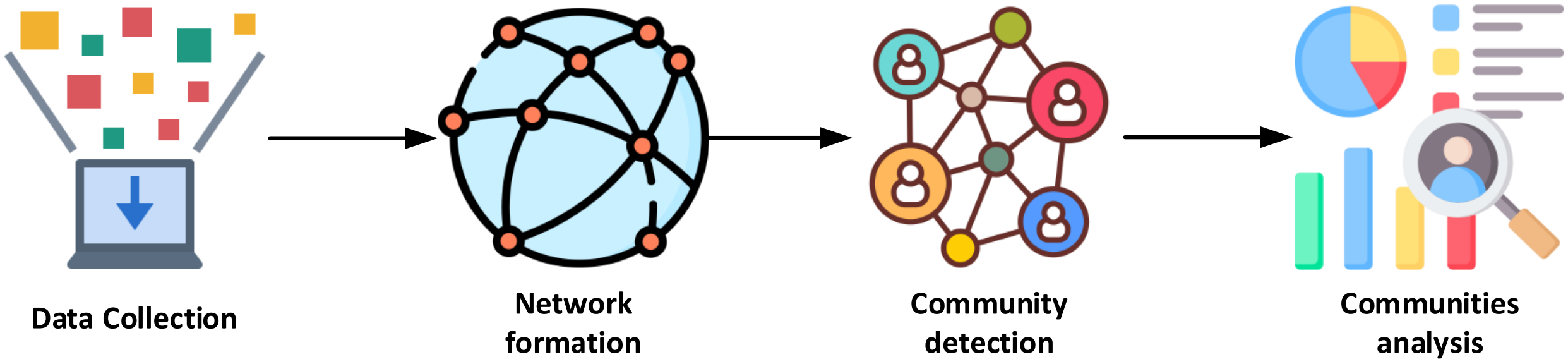

*Figure 3: Research methodology*

In the initial phase, the essential dataset requisite for this investigation is garnered from the renowned "Alexa" website. The data extraction process is facilitated through the utilization of web mining tools and techniques, orchestrated using the Python programming language. Notably, Alexa assumes a pivotal role as a premier platform for global web page ranking and surveillance.

Subsequently, in the second phase, the intricate lattice of Iranian tourism websites is delineated. This is achieved through the deployment of the specialized software "Gephi," a tool dedicated to the intricate analysis of social networks. The outcome is a visual representation of the ecosystem network that encompasses Iranian tourism websites.

The third phase of the study entails an in-depth evaluation of indices pertinent to network analysis. Within this juncture, a suite of standard network analytical parameters is employed, encompassing the following dimensions:

- Graph Size: Denoting the quantity of edges within each discrete graph.
- Node Degree: Quantifying the number of edges converging upon each individual node, thereby expressing its degree.
- Directed Edges: Identifying edges wherein the directionality of the interconnection between nodes holds significance; discerning disparities between the origination and the destination of the linkage.
- Directed Graphs: Encompassing graphs characterized by oriented edges.
- Centrality: Illuminating pivotal nodes within the network, delineating their importance and influence. This pivotal measure serves as a touchstone for network cohesion, unveiling nodes of prominence and impact.

Moreover, the fourth and final phase entails an amalgamation of the visualized network and the indices computed during the third phase, culminating in a comprehensive analysis of the Iranian tourism-oriented website network.

## 4. Findings of the Study

### Step 1: Data Collection

In this phase of the research, using the method of web mining, the most significant Iranian websites related to tourism were found. For targeted sampling in the first step of this phase, initially, a list of the top five hundred Iranian websites was extracted based on their Alexa ranking. Subsequently, the tourism-related websites were manually separated from the rest of the sites. Among the top 500 Iranian websites, were 11 sites associated with the tourism industry, and their names are provided in Table١ .

As part of this research phase, the primary focus was on using web mining techniques to pinpoint the most prominent Iranian websites that pertain to the field of tourism. To achieve this, a systematic approach was adopted. Initially, a compilation of the top 500 Iranian websites was generated based on their Alexa rankings, serving as the foundation for targeted sampling. Subsequently, a meticulous manual curation process was employed to isolate and find websites within this pool that were specifically relevant to the tourism domain. Among the initially found 500 websites, a subset of 11 sites demonstrated a direct affiliation with the tourism industry. Refer to Table١ for a comprehensive list of these noteworthy sites.

Table١ - List of tourism-related websites among the top 500 Iranian websites ranked by Alexa.

| Raw | Website | Alexa Rank |
|---|---|---|
| 1 | Snappfood.ir | 105 |
| 2 | Okala.com | 113 |
| 3 | Alibaba.ir | 178 |
| 4 | Alopeyk.com | 232 |
| 5 | Avval.ir | 339 |
| 6 | Snapp.taxi | 374 |
| 7 | Hamgardi.com | 390 |
| 8 | Snapp.market | 411 |
| 9 | Kojaro.com | 448 |
| 10 | Lastsecond.ir | 463 |
| 11 | Flightio.com | 470 |

During the second phase of this research, an investigation was conducted utilizing the similarity analysis feature, inspired by the mechanism employed by the "Alexa" website. Sites that had survived the filtration process in the initial phase were subjected to this analysis. The similarity analysis is based on a similarity score that reflects the level of resemblance between two websites, considering shared audiences, visitor behavior analysis, and keyword search patterns. Up to 5 sites with distinct similarity scores were identified as most similar to each site under study (*Figure 4*). Notably, sites with higher scores exhibited a more pronounced resemblance to the target site.

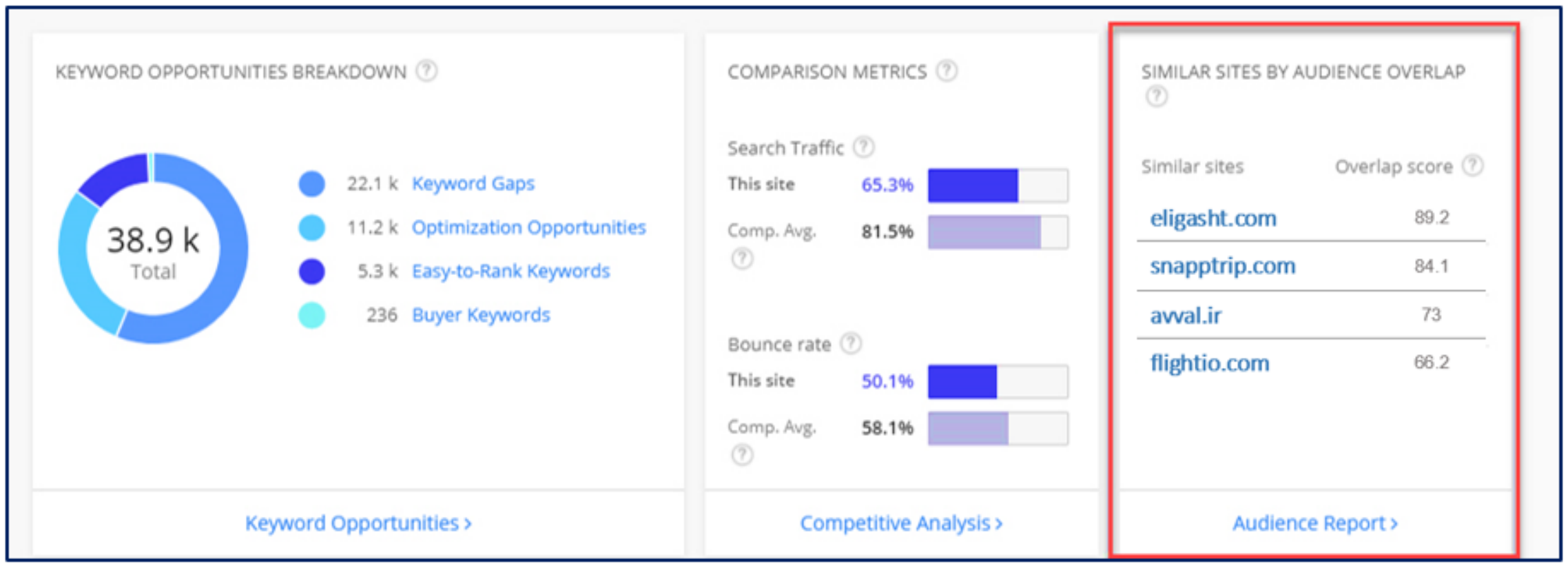

Figure 4- Analyzing the website lastsecond.ir *using* the Alexa website similarity analysis plugin.

Continuing, and using the Python programming language along with the Beautifulsoup library, similar sites to those identified in Step 1 of the research were extracted using the Alexa similarity analysis feature. In this stage, sites with a similarity score of less than 50% were removed due to lower resemblance. Ultimately, 33 similar sites were identified alongside the Step 1 sites.

In Step 3, the Iranian websites related to tourism that were identified in Step 2, referred to as second-layer samples, were subjected to similarity analysis once again using the Alexa website similarity feature. Similar second-layer sites were also extracted. This process was repeated for up to six stages (the reason for terminating this process in the sixth stage was that the extracted cases in the seventh stage and beyond were not only repetitive but also had a similarity score of less than 50%). In all stages, sites with a similarity score exceeding 50% were added to the sample. Many of the sites registered in these six stages had previously been extracted in the first and second stages. Finally, at the conclusion of these six stages, 113 unique sites were recorded.

In the fourth stage, to enhance the research sample, another feature from the "Alexa" website was employed. This feature, termed "Referral Sites," displays the top 5 subsequent and preceding sites that have attracted the highest number of visitors to the studied site (*Figure 5*). Earlier sites visited refer to those that users accessed before visiting tourism-related sites. Furthermore, the subsequent sites pertain to those that users navigate to after visiting tourism-related sites.

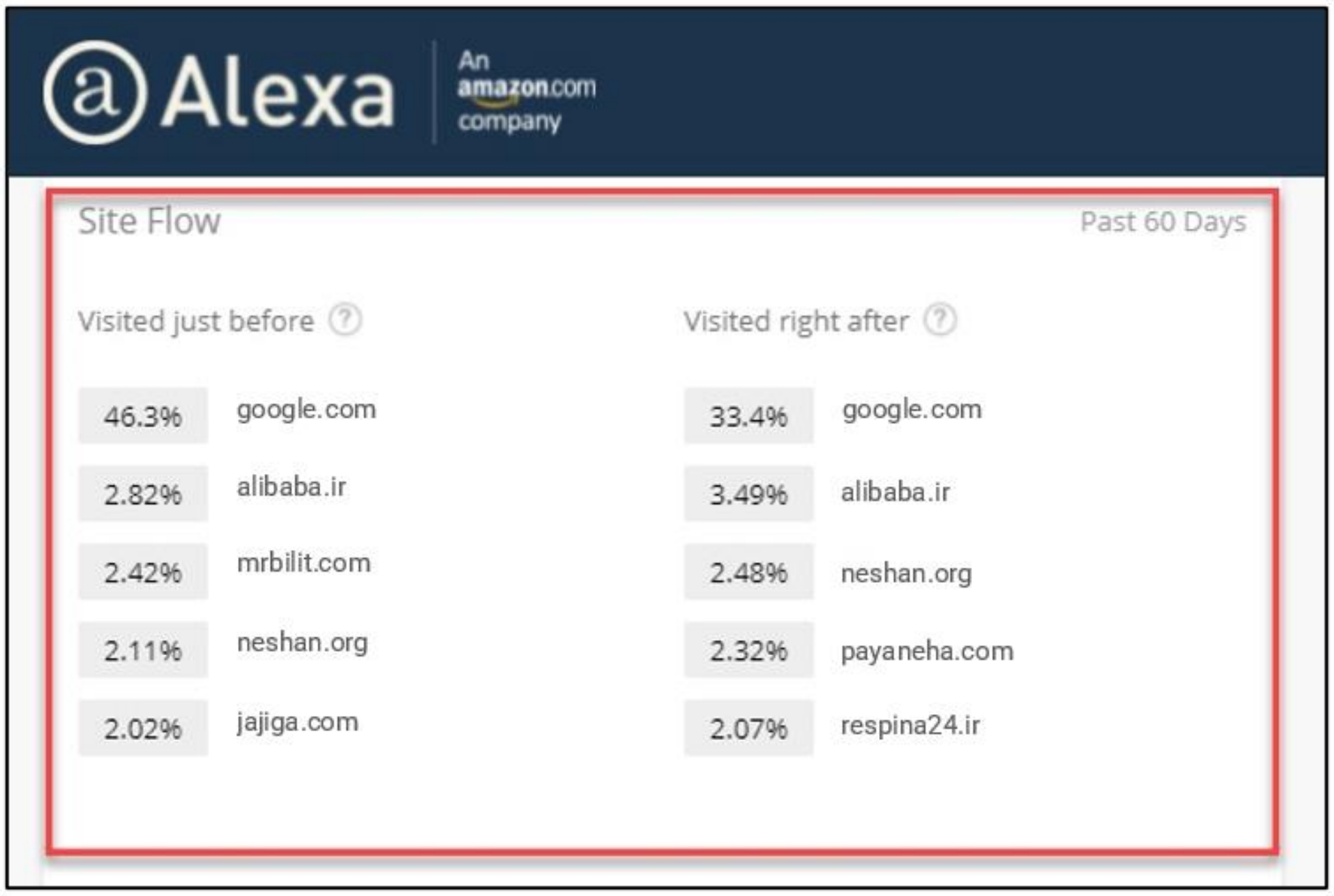

*Figure 5 - Analyzing the website lastseconds.ir using the Alexa website referral sites plugin.*

For each of the extracted instances during the second and third phases, this feature was employed to scrutinize and record up to 5 interactions with other sites as well as up to 5 instances of referring to other sites. Consequently, a dataset of Iranian tourism-related websites was constructed, capturing user interactions. The dataset comprised 498 entities along with the associated connections between them.

**Step 2: Constructing the Web of Interconnected Tourism-Related Websites**

In the second phase, the process involved using the initial dataset formed in the first step. In this step, websites were treated as nodes within a network, and the interactions between users moving between two sites were considered as connections or edges within this network. Consequently, a network of Iranian websites relevant to tourism was constructed. To analyze the network's structural features, a quantitative node weight indicator was employed. To achieve this, user movements from one site to other sites, as well as recorded referrals from other sites to the focal site (considered as a node within the network), were taken into account. The weight attributed to each node was determined based on the number of such incoming and outgoing interactions.

In the subsequent part of this step, irrelevant nodes were eliminated from the network to refine it and render it suitable for more detailed analyses. These nodes had been incorporated into the dataset due to users transitioning from tourism-related sites to unrelated ones. For instance, if users initially accessed a tourism-related site via search engines such as Google or social media platforms like Telegram, Instagram, Twitter, and subsequently navigated to online shopping sites like Digikala.com or other social media, these unrelated sites needed to be excluded from the dataset. Ultimately, after removing extraneous data points, a total of 162 sites aligned with the tourism industry remained in the dataset.

Consistent with the research procedural flow (as depicted in Figure 3), a dataset comprising 162 Iranian websites pertinent to the field of tourism was compiled. This dataset encompasses the websites themselves and the interconnections between them, resulting from user shifts between consecutive sites. Due to the dynamics of user interactions, the social network takes on the form of a directed graph. Furthermore, the weights assigned to the edges (representing connections between two websites) reflect the magnitude of user transitions between those respective sites. The visualization of this interconnected ecosystem of Iranian tourism-related websites can be observed in Figure 6.

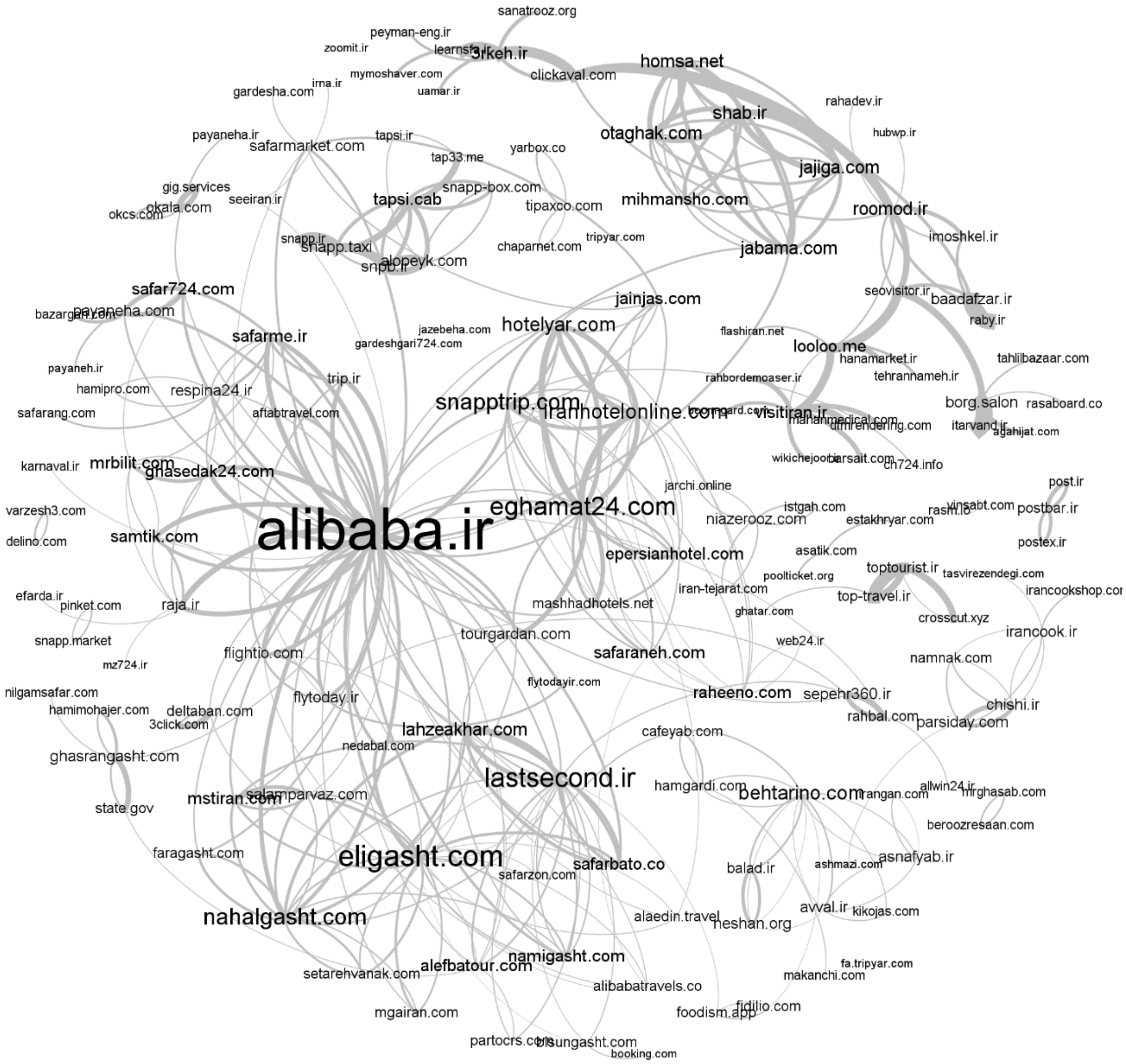

Figure 6 - Tourism-Related Website Ecosystem Analyzed Through User Consumption Behavior

**Step 3: Community Formation within the Ecosystem of Tourism-Related Websites**

To achieve a more nuanced and comprehensive understanding of the tourism ecosystem, it becomes necessary to explore the various groups present within the graph and the interconnectedness among them. The application of community detection techniques within social network analysis proves valuable in uncovering and analyzing these distinctive groups within the network. Essentially, a community comprises members sharing common attributes and collectively influencing the network's behavior. Community detection algorithms categorize nodes with similar attributes, guided by the weights of the edges connecting them within these communities. Subsequently, these detected communities bring together members linked by shared connections. The structure of the network's communities was established using the "Louvain" algorithm, facilitated by the segmentation module of the "Gephi" software tool. After implementing the algorithm and excluding communities with less than 5 members within the network, the outcome yielded 8 discrete communities. These communities encompassed a total of 95 nodes (representing websites) and 306 edges. The composition of these communities is visually depicted in Figure 7.

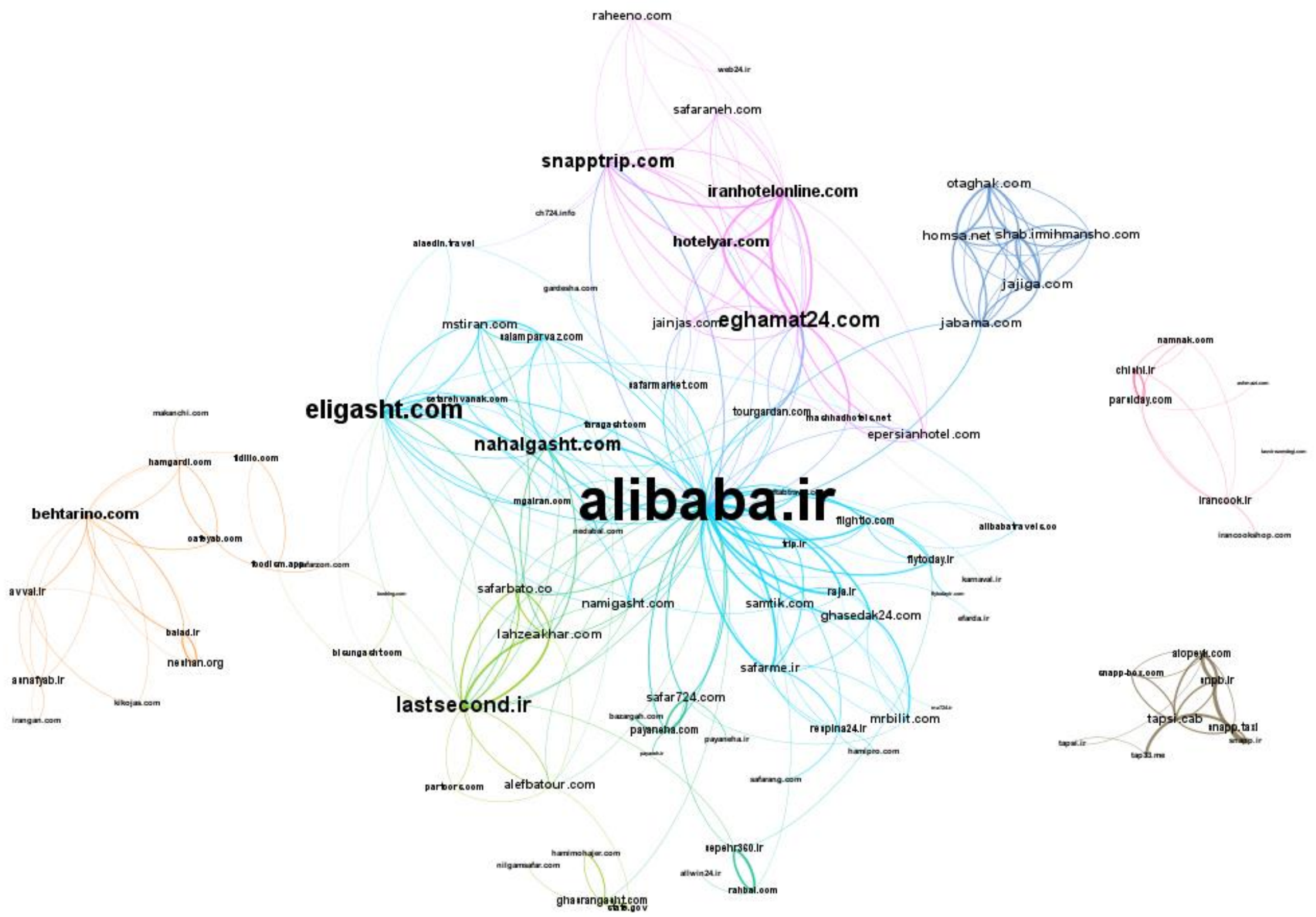

*Figure 7 - Community Formation within the Ecosystem of Tourism-Related Websites Based on User Consumption Behavior*

To facilitate a more detailed analysis of the network, each of the communities within the network has been identified and labeled (Table2). The subsequent analysis will delve into each of the eight identified communities.

*Table2- Community Names within the Ecosystem of Tourism-Related Websites and Their Contribution to the Total Community*

| Raw | Community Name | Percentage of Total | Color |
|---|---|---|---|
| ۱ | Ticket and Tour Booking | ۳۱/۹۶ | Turquoise |
| ۲ | Accommodation (Hotels) | ۱۱/۳۴ | Pink |
| ۳ | Online Taxi Services | ۸/۲۵ | Brown |
| ۴ | Accommodation (Suite and Cottage) | ۶/۱۹ | Purple |
| ۵ | Food and Cooking | ۷/۲۲ | Red |
| ۶ | Location Services | ۱۳/۴ | Orange |
| ۷ | International Tours and Migration | ۱۱/۳۴ | Phosphor Green |
| ۸ | Bus Ticketing | ۸/۲۵ | Bold Green |

1. **Ticket and Tour Booking**: This community accounts for 31.96% of the entire network and is primarily responsible for providing services related to purchasing train, airplane, and bus tickets, as well as hotel reservations and travel tours. Notable sites within this community include "alibaba.ir", " respina24.ir", " eligasht.com", " mrbilit.com" and "flightio.com" The revenue model for these sites is predominantly based on ticket sales, hotel reservations, and travel tour bookings.
2. **Accommodation (Hotels):** The second largest community in the network is comprised of websites offering accommodation services, particularly hotels and apart-hotels. This community contributes 11.34% to the network. Revenue for these sites is generated through intermediary commissions for hotel reservations. Key sites within this community include "snapptrip.com" "hotelyar.com" and "iranhotelonline.com"
3. **Online Taxi Services:** The presence of this community in the network may be attributed to the significant demand for online taxi services during travels. These websites also support intercity travel, contributing to 8% of the Iranian tourism website ecosystem. Notable sites include "snapp.ir" and " tapsi.ir" Revenue for these sites is generated through intermediary commissions.
4. **Accommodation (Suites and Cottage):** This smaller community, constituting approximately 6% of the total network, is distinct for its focus on providing accommodation services in suites, cabins, and small lodgings. Notable sites within this community are "jajiga.com," "otaghak.com," and "jabama.com." Revenue for these sites is earned through rental fees.
5. **Food and Cooking:** This community is likely formed due to tourists' interest in local cuisine. These websites primarily focus on providing culinary education and introducing local dishes. Contributing around 7% to the network, key sites include "ashmazi.com," "irancook.com," and "parsiday.com" The revenue model for these sites mainly revolves around advertising.
6. **Location Services:** The second largest community in the Iranian tourism website ecosystem involves location-based service providers. Comprising 13% of the total network, sites like "Nashan.org," "balad.ir," " cafeyab.com," and "fidilio.com" excel in showcasing locations such as cafes, restaurants, shopping centers, museums, cinemas, and offering map-based services like routing.
7. **International Tours and Migration:** In contrast to the "Ticket and Tour Booking" community, this community focuses more on selling international tours and providing migration services, including visa acquisition and consultation. Representing 11% of the total network, significant sites are "lastsecond.ir" "lahzeakhar.com" and "hamimohajer.com."
8. **Bus Ticketing:** The final community in the Iranian tourism website ecosystem pertains to bus ticketing. This community holds an 8% share in the network. Key sites within this community are "payaneha.com," "payaneh.ir," "bazargah.com," and "safar724.com" These sites generate revenue through ticket sales commissions.

**Step 4: Exploration of the Network Structure in the Ecosystem of Iranian Tourism Websites**

As previously mentioned, the graphical representation is rooted in user transitions from one website to another. Consequently, the graph depicted in Figure 7 is directed in nature. A directed network permits the examination of user movement trajectories between various communities. Thus, comparing the inflows and outflows of a given node, along with pre-and post-connection dynamics within the network, holds substantial significance and enables an understanding of the ecosystem's intricate dynamics based on user behavior. To accomplish this, each community is initially treated as a unified node. This approach implies that instead of analyzing individual components within each community, all members are collectively regarded as a singular entity, and the interactions among them are subject to analysis. The weight of the connection between any two nodes (referred to as edge weight) indicates the average weight of interactions from members of the source community to members of the target community. This weight is calculated using the subsequent formula:

$$W_{ij} = \frac{\sum_{i=1}^{n} \sum_{j=1}^{m} W_{ij}}{n * m}$$

Table 3 elaborates on the intricate connections between different communities. The values within this table signify the average weight of interactions between all members of the source community and all members of the target community. Each row in the table represents the source community, while the columns illustrate the destination of the connections. Given the directed nature of the communication network in the ecosystem, the values within the table are asymmetric, and vacant cells in the table denote the absence of interactions between members of distinct communities. The color coding employed in this table ranges from green (indicating higher values) to red (indicating lower values). The maximum value within the table is 2.4 (highlighted in bold green), while the minimum value is 1.3 (highlighted in bold red).

*Table 3- Distribution of Weighted Inter-Community Connections within the Ecosystem of Tourism Websites*

| From - To | Ticket and Tour Booking | Accommodation (Hotels) | Location Services | Online Taxi Services | International Tours and Migration | Accommodation Suite and) (Cottage | Food and Cooking | Bus Ticketing |
|---|---|---|---|---|---|---|---|---|
| Ticket and Tour Booking | | 25/6 | 1/5 | | 45/1 | 4/2 | | 15/2 |
| Accommodation (Hotels) | 34/7 | | | | | | | |
| Location Services | | | | | 1/2 | | | |
| Online Taxi Services | | | | | | | | |
| International Tours and Migration | 30/2 | | 1/2 | | | | | 2/4 |
| Accommodation (Suite and Cottage) | 4/8 | | | | | | | |
| Food and Cooking | | | | | | | | |
| Bus Ticketing | 11/6 | | | | | | | |

Drawing insights from the findings of Table 3, Figure 8 visualizes the inter-community relationships.

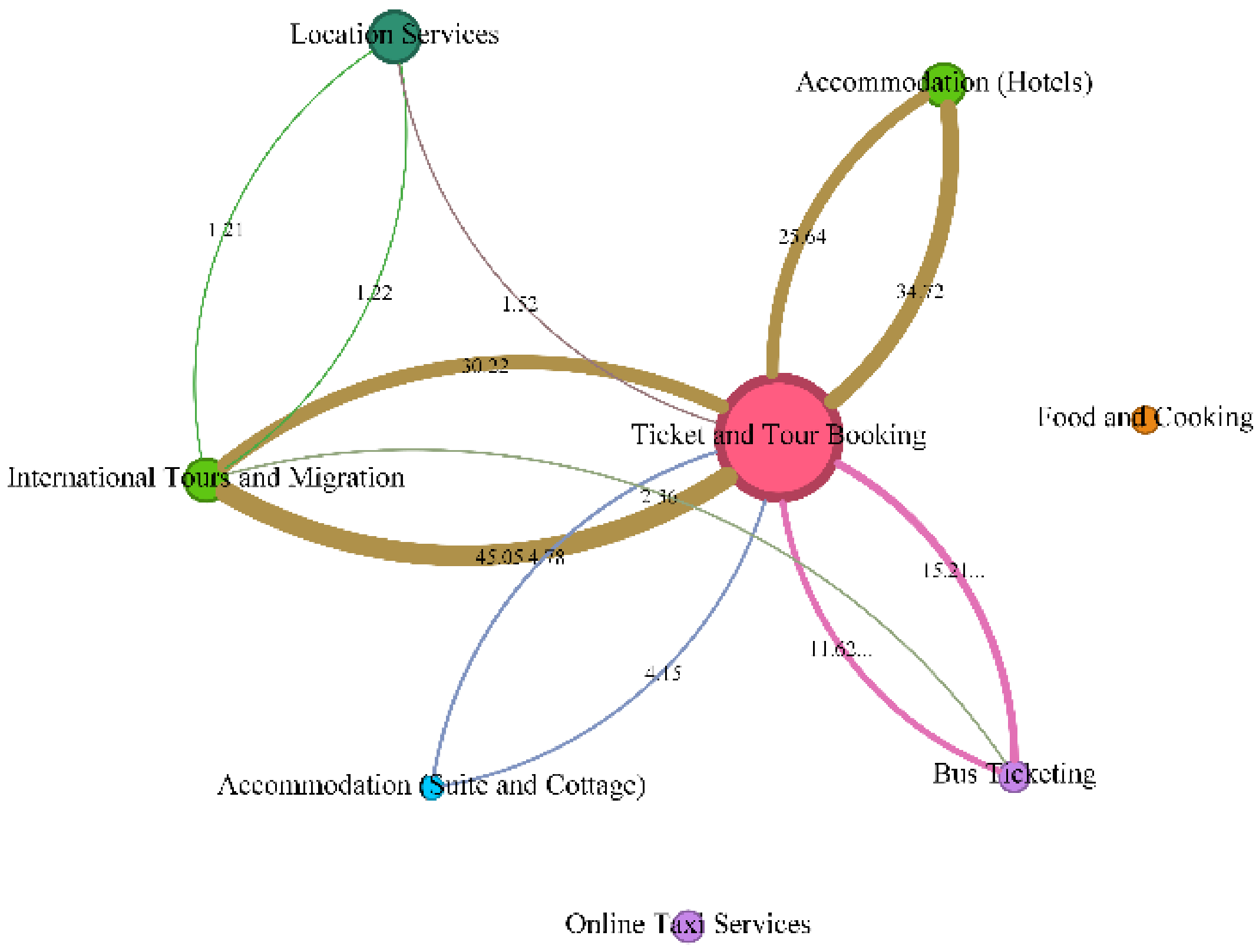

Figure 8 - Network Visualization of Inter-Community Connections within the Ecosystem of Iranian Tourism Websites

This graph comprises 8 nodes and 12 edges. Each node represents the communities uncovered within the ecosystem of tourism websites, and the edges symbolize the average weight of interactions from members of the source community to members of the target community. The numeric labels on each edge denote the corresponding connection weights. The size of the nodes corresponds to their weights; larger nodes denote greater numbers of connections in the graph. For example, the "Ticket and Tour Booking" community, positioned at the graph's center and exhibiting a larger size than other nodes, is the most prominent community within the ecosystem. Notably, the graph is directed, signifying the directional flow of users from the source community to the destination community. Hence, employing this graph and considering the weight of connections between various communities allows for an insightful analysis of user behavior within the Iranian tourism website ecosystem.

## 5. Conclusion

The objective of this study was to identify, categorize, and analyze the ecosystem of Iranian tourism websites based on user behavior. To achieve this, the ecosystem was analyzed using social network analysis, focusing on the connections between websites through user movements. It can be stated that attributing authenticity to user behavior is one of the ways to comprehend the media ecosystem, and in this study, it was achieved by exploring the similarities and differences in user behavior across websites. These variations and similarities led to the formation of distinct communities. These communities consist of the most similar elements to each other and possess the greatest differences from other communities. Members within a community are so alike that they can be regarded as a unified whole based on shared characteristics. According to the research results, the ecosystem of Iranian tourism websites comprises 95 core members. The separation of communities based on user behavior forms networks that are shared among community members while also differentiating them from other communities. Eight website communities within the tourism sector were identified based on user behavior: "Ticket and Tour Booking," "Accommodation (Hotels)," "Location Services," "Online Taxi Services," "Foreign Tours and Migration," "Accommodation (Suites and Cottages)," "Food and Cooking," and "Bus Ticket Booking." Subsequently, the relationships between these clusters were examined, and the strengths and weaknesses of their connections were identified.

*Table 4- Inter-Community Relationships*

| *Community* | *Strong Connections* | *Weak Connections* |
|---|---|---|
| *Ticket and Tour Booking* | International Tours and Migration (45.1), Accommodation (Hotels) (25.6), Bus Ticketing (15.2) | Location Services (1.5), Accommodation (Suite and Cottage) (4.2) |
| *Accommodation (Hotels)* | Ticket and Tour Booking (34.7) | |
| *Online Taxi Services* | No Connections | |
| *Accommodation (Suite and Cottage)* | | Ticket and Tour Booking (4.8) |
| *Food and Cooking* | No Connections | |
| *Location Services* | | International Tours and Migration (1.2), Bus Ticketing (2.4) |
| *International Tours and Migration* | Ticket and Tour Booking (30.2) | Location Services (1.2) |
| *Bus Ticketing* | Ticket and Tour Booking (11.6) | |

**Analyzing Inter-Community Relationships**

1. **Ticket and Tour Booking:** A strong connection between this cluster and "Accommodation (Hotels)" can be attributed to the complementary nature of these services. When individuals plan trips, they often require transportation and lodging. Similarly, the strong connection with the "Foreign Tours and Migration" community indicates a synergy between these two services, as travelers may seek organized tours when visiting foreign destinations. The weaker connections to "Location Services" and "Bus Ticket Booking" suggest that these communities might not be perceived as essential components of the overall travel experience or might be easily accessible through alternative methods.

2. **Accommodation (Hotels):** The strong connection with "Ticket and Tour Booking" implies that travelers prioritize finalizing their accommodation after planning their trips. The absence of connections to other clusters may indicate that hotel reservations are typically considered separately from other travel-related services.
3. **Foreign Tours and Migration:** The importance of the strong connection between this community and "Ticket and Tour Booking" is evident. This behavior could arise from price and cost comparisons across different websites. Additionally, due to the comprehensive services offered through tours, the lack of connections to other communities appears logical, as these services might serve distinct purposes.
4. **Accommodation (Suites and Cottages):** In contrast to the "Accommodation (Hotels)" community, which focuses on hotel reservations, this community holds lower weight in its connections to the broader network. Hence, it can be deduced that online reservations for suites and cottages have not gained substantial traction within the country's tourism industry to date. This presents a significant untapped potential for growth in the sector.
5. **Location Services:** The limited connection with the "Foreign Tours and Migration" community indicates that tourists might resort to alternative sources, such as specialized travel agencies or foreign websites, to gather information about foreign destinations. Moreover, weaker links to other clusters might imply that location and mapping services are either less utilized or integrated into the broader travel planning process.
6. **Bus Ticket Booking:** Given the relatively strong connection to the "Ticket and Tour Booking" community, a strategy to integrate these websites could be considered. This could result in a favorable synergy within the tourism industry, providing customers with a comprehensive travel solution and maximizing convenience, satisfaction, and competition.
7. **Online Taxi Services and Food/Cooking:** The absence of connections to other clusters suggests that these services might not be perceived as integral parts of the travel experience or might have limited demand among tourists.

The current research demonstrates that the needs of users have a significant influence on their interaction with tourism websites and even on their transitions between different websites. Furthermore, it can be argued that the role of the user has taken on a more exploratory dimension, with users actively seeking to address their needs by navigating various communities. This study represents the initial step toward comprehensively analyzing the tourism ecosystem of the country. By complementing other research efforts that delve into various influencing factors on the tourism ecosystem, such as the inherent characteristics of websites, policies implemented by decision-makers, content limitations, technical infrastructure, and cultural attributes, more comprehensive insights can be attained.

Analyzing communities based on user behaviors and flows yielded insightful results. Each stakeholder within Iran's tourism ecosystem can gain a unique perspective on the identified ecosystem, tailored to their respective concerns. Following this, some policy recommendations for stakeholders within the industry—namely, service providers, investors, and legislators—are outlined:

1. **Strengthen Connections with Bus Ticket Sales:** Identifying opportunities to enhance the connection between the "Ticket and Tour Booking" cluster and bus ticket sales could potentially be achieved through joint marketing efforts or integrated reservation systems.
2. **Enhance Public-Private Partnerships:** Encouraging collaboration between accommodation providers, tour operators, and transportation companies to elevate the overall tourist experience.

3. **Promote Ecotourism Culture:** Supporting and developing indigenous accommodations like cottages and suites with an emphasis on introducing local tourism opportunities and fostering local tourism economies.
4. **Culinary Tourism Promotion:** Introducing initiatives to highlight culinary education and gastronomy, such as food festivals or culinary tours, to attract tourists interested in exploring Iranian cuisine and culinary traditions.
5. **Encourage Innovation in Online Taxi Services:** Supporting technological advancements and innovation in online taxi services to provide efficient and convenient transportation options for tourists within Iran.
6. **Improve Timely Access to Location and Mapping Services:** Investing in improving visibility and usability of location-based services and mapping to ensure tourists have accurate and reliable information about destinations within Iran.

## references

Ali Ghasemian, S., Rahil, K., & Mohammad, A. (2021). A New Approach in Marketing Research: Identifying the Customer Expected Value through Machine Learning and Big Data Analysis in the Tourism Industry. *Asia-Pacific Journal of Management and Technology (AJMT)*, *2*(3), 26-42. https://doi.org/10.46977/apjmt.2022v02i03.004

Arroyo, C. G., Knollenberg, W., & Barbieri, C. (2021). Inputs and outputs of craft beverage tourism: The Destination Resources Acceleration Framework. *Annals of Tourism Research*, *86*, 103102.

Bedi, P., & Sharma, C. (2016). Community detection in social networks. *Wiley Interdisciplinary Reviews: Data Mining and Knowledge Discovery*, *6*(3), 115-135.

Beladi, H., Chao, C.-C., Ee, M. S., & Hollas, D. (2017). Does Medical Tourism Promote Economic Growth? A Cross-Country Analysis. *Journal of Travel Research*, *58*(1), 121-135. https://doi.org/10.1177/0047287517735909

Bodhanwala, S., & Bodhanwala, R. (2022). Exploring relationship between sustainability and firm performance in travel and tourism industry: a global evidence. *Social Responsibility Journal*, *18*(7), 1251-1269.

Büyüközkan, G., & Ergün, B. (2011). Intelligent system applications in electronic tourism. *Expert Systems with Applications*, *38*(6), 6586-6598.

Carvalho, P., & Alves, H. (2023). Customer value co-creation in the hospitality and tourism industry: a systematic literature review. *International Journal of Contemporary Hospitality Management*, *35*(1), 250-273.

Chen, Y., Chen, R., Hou, J., Hou, M., & Xie, X. (2021). Research on users' participation mechanisms in virtual tourism communities by Bayesian network. *Knowledge-Based Systems*, *226*, 107161.

Christou, P., Hadjielias, E., Simillidou, A., & Kvasova, O. (2023). The use of intelligent automation as a form of digital transformation in tourism: Towards a hybrid experiential offering. *Journal of Business Research*, *155*, 113415.

Dolnicar, S., Grün, B., Leisch, F., & Schmidt, K. (2014). Required sample sizes for data-driven market segmentation analyses in tourism. *Journal of Travel Research*, *53*(3), 296-306.

Fararni, K. A., Nafis, F., Aghoutane, B., Yahyaouy, A., Riffi, J., & Sabri, A. (2021). Hybrid recommender system for tourism based on big data and AI: A conceptual framework. *Big Data Mining and Analytics*, *4*(1), 47-55. https://doi.org/10.26599/BDMA.2020.9020015

Gudkov, A., & Dedkova, E. (2020). Development and financial support of tourism exports in the digital economy. *Journal of Digital Science*, *2*(1), 54-66.

Gupta, G. (2019). Inclusive use of digital marketing in tourism industry. Information Systems Design and Intelligent Applications: Proceedings of Fifth International Conference INDIA 2018 Volume 1,

Gupta, S., Modgil, S., Lee, C.-K., & Sivarajah, U. (2023). The future is yesterday: Use of AI-driven facial recognition to enhance value in the travel and tourism industry. *Information Systems Frontiers*, *25*(3), 1179-1195. https://doi.org/10.1007/s10796-022-10271-8

Gursoy, D., Malodia, S., & Dhir, A. (2022). The metaverse in the hospitality and tourism industry: An overview of current trends and future research directions. *Journal of Hospitality Marketing & Management*, *31*(5), 527-534.

HabibAgahi, M. R., Kermani, M. A. M. A., & Maghsoudi, M. (2022). On the Co-authorship network analysis in the Process Mining research Community: A social network analysis perspective. *Expert Systems with Applications*, *206*, 117853.

Holloway, J. C., & Humphreys, C. (2022). *The business of tourism*. Sage.

Jalilvand Khosravi, M., Maghsoudi, M., & Salavatian, S. (2022). Identifying and Clustering Users of VOD Platforms Using SNA Technique: A case study of Cinemamarket. *New Marketing Research Journal*, *11*(4), 20-21.

Jeaheng, Y., & Han, H. (2020). Thai street food in the fast growing global food tourism industry: Preference and behaviors of food tourists. *Journal of Hospitality and Tourism Management*, *45*, 641-655.

Kermani, M. A., Maghsoudi, M., Hamedani, M. S., & Bozorgipour, A. (2022). Analyzing the interorganizational collaborations in crisis management in coping with COVID-19 using social network analysis: Case of Iran. *Journal of emergency management (Weston, Mass.)*, *20*(3), 249-266.

Kermani, M. A. M. A., Maghsoudi, M., Hamedani, M. S., & Bozorgipour, A. (2022). Analyzing the interorganizational collaborations in crisis management in coping with COVID-19 using social network analysis: Case of Iran. *Journal of emergency management (Weston, Mass.)*, *20*(3), 249-266.

Kozak, M., & Buhalis, D. (2019). Cross–border tourism destination marketing: Prerequisites and critical success factors. *Journal of Destination Marketing & Management*, *14*, 100392.

Kwok, A. O. (2023). The next frontier of the Internet of Behaviors: data-driven nudging in smart tourism. *Journal of Tourism Futures*.

Maghsoudi, M., Jalilvand Khosravi, M., & Salavatian, S. (2023). Analyzing the Ecosystem of Iranian websites for Watching and Downloading videos based on User Behavior. *Iranian Journal of Information Processing and Management*, *38*(3), 1067-1094.

Maghsoudi, M., & Nezafati, N. (2023). Navigating the Acceptance of Implementing Business Intelligence in Organizations: A System Dynamics Approach. *Telematics and Informatics Reports*, 100070. https://doi.org/https://doi.org/10.1016/j.teler.2023.100070

Maghsoudi, M., Shokouhyar, S., Khanizadeh, S., & Shokoohyar, S. (2023). Towards a taxonomy of waste management research: An application of community detection in keyword network. *Journal of cleaner production*, 136587. https://doi.org/https://doi.org/10.1016/j.jclepro.2023.136587

Maghsoudi, M., & Shumaly, S. Study of co-authorship network of papers in all sessions of IIIEC Conference.

Martín Martín, J. M., Prados-Castillo, J. F., de Castro-Pardo, M., & Jimenez Aguilera, J. D. D. (2021). Exploring conflicts between stakeholders in tourism industry. Citizen attitude toward peer-to-peer accommodation platforms. *International Journal of Conflict Management*, *32*(4), 697-721.

Newman, M. E. (2006). Modularity and community structure in networks. *Proceedings of the national academy of sciences*, *103*(23), 8577-8582.

Obogo, J. U., & Adedoyin, F. F. (2021). Data-Driven Business Analytics for the Tourism Industry in the UK: A Machine Learning Experiment Post-COVID. 2021 IEEE 23rd Conference on Business Informatics (CBI),

Okumus, B. (2021). Food tourism research: a perspective article. *Tourism Review*, *76*(1), 38-42.

Ozdemir, O., Dogru, T., Kizildag, M., & Erkmen, E. (2023). A critical reflection on digitalization for the hospitality and tourism industry: value implications for stakeholders. *International Journal of Contemporary Hospitality Management*.

Park, E., Park, J., & Hu, M. (2021). Tourism demand forecasting with online news data mining. *Annals of Tourism Research*, *90*, 103273. https://doi.org/https://doi.org/10.1016/j.annals.2021.103273

Pasca, M. G., Renzi, M. F., Di Pietro, L., & Guglielmetti Mugion, R. (2021). Gamification in tourism and hospitality research in the era of digital platforms: a systematic literature review. *Journal of Service Theory and Practice*, *31*(5), 691-737.

Polese, F., Botti, A., & Monda, A. (2022). Value co-creation and data-driven orientation: reflections on restaurant management practices during COVID-19 in Italy. *Transforming Government: People, Process and Policy*, *16*(2), 172-184.

Ridzuan, M. R., & Abd Rahman, N. A. S. (2021). The deployment of fiscal policy in several ASEAN countries in dampening the impact of COVID-19. *Journal of Emerging Economies and Islamic Research*, *9*(1), 16-28.

Rwigema, P. (2021). Critical analysis of tourism industry and sustainable development during Covid-19 pandemic. A case of Rwanda. *The Strategic Journal of Business & Change Management*, *8*(4), 594.

Shapoval, V., Wang, M. C., Hara, T., & Shioya, H. (2018). Data Mining in Tourism Data Analysis: Inbound Visitors to Japan. *Journal of Travel Research*, *57*(3), 310-323. https://doi.org/10.1177/0047287517696960

Weaver, A. (2021). Tourism, big data, and a crisis of analysis. *Annals of Tourism Research*, *88*, 103158. https://doi.org/https://doi.org/10.1016/j.annals.2021.103158

Yanes, A., Zielinski, S., Diaz Cano, M., & Kim, S.-i. (2019). Community-based tourism in developing countries: A framework for policy evaluation. *Sustainability*, *11*(9), 2506.

Zafarani, R., Abbasi, M. A., & Liu, H. (2014). *Social media mining: an introduction*. Cambridge University Press.

Zhang, J., Wu, T., & Fan, Z. (2019). Research on precision marketing model of tourism industry based on user's mobile behavior trajectory. *Mobile Information Systems*, *2019*.

Zhao, X., & Yu, G. (2021). Data-Driven Spatial Econometric Analysis Model for Regional Tourism Development. *Mathematical Problems in Engineering*, *2021*, 6631833. https://doi.org/10.1155/2021/6631833

Zohdi, M., Maghsoudi, M., & Nooralizadeh, H. (2022). Providing a User-Based Behavior Model to Recommend a Movie Using the Social Network Analysis (Case Study: CinemaMarket). *Sciences and Techniques of Information Management*, *8*(1), 451-484.